\documentclass[aps,prl,reprint,superscriptaddress,showpacs]{revtex4-1}

\usepackage{graphicx}
\begin{document}

%\preprint{}

\title{Enhanced $\gamma$-Ray Emission from Neutron Unbound States Populated in $\beta$ Decay}


%\author{}
%\thanks{}

\author{J. L. Tain}
\email[Corresponding author:]{tain@ific.uv.es}
\author{E. Valencia}
\author{A. Algora}
\author{J. Agramunt}
\author{B. Rubio}
\affiliation{Instituto de Fisica Corpuscular (CSIC-Universitat de Valencia),
  Apdo. Correos 22085, E-46071 Valencia, Spain}

\author{S. Rice}
\author{W. Gelletly}
\author{P. Regan}
\affiliation{University of Surrey, Department of Physics, Guildford GU2 7XH, United Kingdom}

\author{A.-A. Zakari-Issoufou}
\author{M. Fallot}
\author{A. Porta}
\affiliation{SUBATECH, CNRS/IN2P3, Universit\'e de Nantes, Ecole des Mines, F-44307 Nantes, France}

\author{J. Rissanen}
\author{T. Eronen}
\affiliation{University of Jyv\"askyl\"a, Department of Physics, P.O. Box 35, FI-40014 Jyv\"askyl\"a, Finland}

\author{J. \"Ayst\"o}
\affiliation{Helsinki Institute of Physics, FI-00014 University of Helsinki, Finland}

\author{L. Batist}
\affiliation{Petersburg Nuclear Physics Institute, RU-188300 Gatchina, Russia}

\author{M. Bowry}
\affiliation{University of Surrey, Department of Physics, Guildford GU2 7XH, United Kingdom}

\author{V. M. Bui}
\affiliation{SUBATECH, CNRS/IN2P3, Universit\'e de Nantes, Ecole des Mines, F-44307 Nantes, France}

\author{R. Caballero-Folch}
\affiliation{Universitat Politecnica de Catalunya, E-08028 Barcelona, Spain}

\author{D. Cano-Ott}
\affiliation{Centro de Investigaciones Energ\'eticas Medioambientales y
  Tecn\'ologicas, E-28040 Madrid, Spain}

\author{V.-V. Elomaa}
\affiliation{University of Jyv\"askyl\"a, Department of Physics, P.O. Box 35, FI-40014 Jyv\"askyl\"a, Finland}

\author{E. Estevez}
\affiliation{Instituto de Fisica Corpuscular (CSIC-Universitat de Valencia),
  Apdo. Correos 22085, E-46071 Valencia, Spain}

\author{G. F. Farrelly}
\affiliation{University of Surrey, Department of Physics, Guildford GU2 7XH, United Kingdom}

\author{A. R. Garcia}
\affiliation{Centro de Investigaciones Energ\'eticas Medioambientales y
  Tecn\'ologicas, E-28040 Madrid, Spain}

\author{B. Gomez-Hornillos}
\author{V. Gorlychev}
\affiliation{Universitat Politecnica de Catalunya, E-08028 Barcelona, Spain}

\author{J. Hakala}
\affiliation{University of Jyv\"askyl\"a, Department of Physics, P.O. Box 35, FI-40014 Jyv\"askyl\"a, Finland}

\author{M.D. Jordan}
\affiliation{Instituto de Fisica Corpuscular (CSIC-Universitat de Valencia),
  Apdo. Correos 22085, E-46071 Valencia, Spain}

\author{A. Jokinen}
\author{V. S. Kolhinen}
\affiliation{University of Jyv\"askyl\"a, Department of Physics, P.O. Box 35, FI-40014 Jyv\"askyl\"a, Finland}

\author{F. G. Kondev}
\affiliation{Nuclear Engineering Division, Argonne National Laboratory, 
Argonne, Illinois 60439, USA}

\author{T. Mart\'{\i}nez}
\author{E. Mendoza}
\affiliation{Centro de Investigaciones Energ\'eticas Medioambientales y
  Tecn\'ologicas, E-28040 Madrid, Spain}

\author{I. Moore}
\author{H. Penttil\"a}
\affiliation{University of Jyv\"askyl\"a, Department of Physics, P.O. Box 35, FI-40014 Jyv\"askyl\"a, Finland}

\author{Zs. Podoly\'ak}
\affiliation{University of Surrey, Department of Physics, Guildford GU2 7XH, United Kingdom}

\author{M. Reponen}
\author{V. Sonnenschein}
\affiliation{University of Jyv\"askyl\"a, Department of Physics, P.O. Box 35, FI-40014 Jyv\"askyl\"a, Finland}

\author{A. A. Sonzogni}
\affiliation{NNDC, Brookhaven National Laboratory, Upton, New York 11973, USA}


\date{\today}

\begin{abstract}
Total absorption spectroscopy was used to investigate
the $\beta$-decay intensity to states above the neutron
separation energy followed by $\gamma$-ray emission in $^{87,88}$Br and $^{94}$Rb.
Accurate results were obtained thanks to a careful control of systematic errors.
An unexpectedly large $\gamma$ intensity was observed in all three cases extending well beyond
the excitation energy region where neutron penetration is hindered by low neutron energy. 
The $\gamma$ branching as a function of excitation energy
was compared to Hauser-Feshbach model calculations.
For $^{87}$Br and $^{88}$Br the $\gamma$ branching reaches 57\% and 20\% respectively,
and could be explained as a nuclear structure effect. Some of the
states populated in the daughter can only decay through the emission of a large
orbital angular momentum neutron with a strongly reduced barrier penetrability.
In the case of neutron-rich $^{94}$Rb
the observed 4.5\% branching is
much larger than the calculations performed with standard nuclear statistical 
model parameters, even after proper correction for fluctuation effects on individual transition widths.
The difference can be reconciled  introducing an enhancement of one order-of-magnitude
in the photon strength to neutron strength ratio. 
An increase in the photon strength function of such magnitude for very neutron-rich nuclei,
if it proved to be correct, leads to a similar increase in the
$(\mathrm{n},\gamma)$ cross section that would have 
an impact on $r$ process abundance calculations.

\end{abstract}

%\keywords{}

\maketitle

Neutron unbound states can be populated in the $\beta$ decay of
very neutron-rich nuclei, when the neutron separation energy
$S_{n}$ in the daughter nucleus is lower than the decay energy window $Q_{\beta}$.
Given the relative strengths of strong and electromagnetic interactions
these states decay preferentially by neutron emission.
Beta delayed $\gamma$-ray emission from states above $S_{n}$
was first observed in 1972 in the decay of $^{87}$Br~\cite{sla72}. 
Since then it has been observed in a handful of cases:
$^{137}$I~\cite{nuh75}, $^{93}$Rb~\cite{bis77}, $^{85}$As~\cite{kra79}, 
$^{141}$Cs~\cite{yam82}, $^{95}$Rb~\cite{kra83a}, and $^{94}$Rb~\cite{alk89}. 
The paucity of information is related to the difficulty of detecting
weak high-energy $\gamma$-ray cascades 
with the germanium detectors that are usually employed in $\beta$-decay studies.
This problem has become known
as the {\it Pandemonium} effect~\cite{har77} and it also affects the accuracy of the data. 

There is an analogy~\cite{kra83b} between this decay process and 
neutron capture reactions which populate 
states in the compound nucleus that re-emit a neutron (elastic channel) or
de-excite by $\gamma$ rays (radiative capture). 
Indeed the reaction
cross section is parametrized in terms of neutron and $\gamma$ widths,
$\Gamma_{n}$ and $\Gamma_{\gamma}$ respectively,
which also determines the fraction of $\beta$ intensity above $S_{n}$ 
that proceeds by neutron or $\gamma$ emission.
Radiative capture $(\mathrm{n},\gamma)$
cross sections for very neutron-rich nuclei are 
a key ingredient in reaction network calculations used to obtain
the yield of elements heavier than iron in the rapid ($r$) neutron capture
process occurring in explosive-like stellar events. It has been shown~\cite{gor98,sur01,arc11}
that the abundance distributions in different astrophysical scenarios
are sensitive to $(\mathrm{n},\gamma)$ cross sections. 
In the classical ``hot'' $r$ process late captures during freeze-out
modify the final element abundance. In the ``cold" $r$ process
the competition between neutron captures and $\beta$ decays
determines the formation path.
Cross section values for these exotic nuclei
are taken from Hauser-Feshbach model calculations~\cite{rau00}, 
which are based on a few quantities describing average nuclear properties: 
nuclear level densities (NLD), photon strength
functions (PSF) and neutron transmission coefficients (NTC).
Since these quantities are adjusted to experiment close to $\beta$ stability
it is crucial to find means to verify the predictions for very
neutron-rich nuclei. 

The Total Absorption Gamma-ray Spectroscopy (TAGS) technique
aims at detecting cascades rather than individual $\gamma$ rays
using large $4\pi$ scintillation detectors.
The superiority of this method over high-resolution germanium spectroscopy
to locate missing $\beta$ intensity has been
demonstrated before~\cite{alg99,hu99}. However its application in the present case is 
very challenging, since the expected $\gamma$-branching is very small
and located at rather high excitation energies. 
As a matter of fact previous attempts at LNPI~\cite{alk89} with a similar aim 
did not lead to clear conclusions. 
In this Letter we propose and demonstrate for the first time the use of the 
TAGS technique
to study $\gamma$-ray emission above $S_{n}$ in
$\beta$-delayed neutron emitters and extract accurate information
that can be used to improve $(\mathrm{n},\gamma)$ 
cross section estimates far from $\beta$ stability.

Neutron capture and transmission reactions have been extensively used~\cite{mug06}
to determine neutron and $\gamma$ widths (or related strength functions).
An inspection of Ref.~\cite{mug06} shows
that in general $\Gamma_{n}$ is orders-of-magnitude larger
than $\Gamma_{\gamma}$.
In the decay of $^{87}$Br, which is the best studied case~\cite{sla72,tov75,nuh77,ram83},
a dozen states emitting single $\gamma$ rays 
were identified within 250 keV above $S_{n}$
collecting about 0.5\% of the decay intensity to be compared
with a neutron emission probability of 2.6\%. 
The observation of such relatively high $\gamma$-ray intensity was
explained as being due to a nuclear structure effect:
some of the levels populated can only decay by emission
of neutrons with large orbital angular momentum $l$,
which is strongly hindered.
In addition it has been pointed out~\cite{jon76} that a sizable $\gamma$-ray
emission from neutron unbound states can be a manifestation of Porter-Thomas (PT)
statistical fluctuations in the strength of individual transitions. 
The role and relative importance of both mechanisms should be
investigated.

We present here the results of measurements 
for three known
neutron emitters, $^{87}$Br~\cite{br87}, $^{88}$Br~\cite{br88}  and $^{94}$Rb~\cite{rb94} , 
using a newly developed TAGS spectrometer.
The results for $^{93}$Rb, also measured, will be presented later~\cite{zak15}.
The measurements were performed at the  IGISOL mass separator~\cite{ays01}
of the University of Jyv\"askyl\"a. 
The isotopes were produced by proton-induced fission of uranium and 
the mass-separated beam was cleaned from isobaric
contamination using the JYFLTRAP Penning trap~\cite{kol04,ero12}. 
The resulting beam 
was implanted at the centre of the spectrometer
onto a movable tape which periodically removed the activity
to minimize daughter contamination. 
Behind the tape was placed a 0.5~mm thick Si detector with a $\beta$-detection efficiency 
of about 30\%. 
The Valencia-Surrey Total Absorption Spectrometer {\it Rocinante} 
is a cylindrical 12-fold
segmented BaF$_{2}$ detector with a length and external diameter of 25~cm, 
and a longitudinal hole of 5~cm diameter.
The detection efficiency for single $\gamma$ rays is larger than 80\%.
The spectrometer has a reduced neutron sensitivity in comparison to NaI(Tl)
detectors, a key feature in the present application. It also
allows the measurement of multiplicities which helps in the data analysis.
In order to eliminate the detector intrinsic background 
and the ambient background 
we use $\beta$-gated TAGS spectra
in the present analysis. Nevertheless other sources of
spectrum contamination need to be characterized accurately. 

In the first place the decay descendant contamination,
was computed using the Geant4 simulation toolkit~\cite{geant4}. 
In the case of the daughter decay we use an event generator based on  
the well known decay level scheme~\cite{br87,br88,rb94}. 
The calculated normalization factor was adjusted to provide the best fit to
the recorded spectrum.
The measurement of $^{88}$Br was accidentally
contaminated by $^{94}$Y, the long-lived grand-daughter of $^{94}$Rb,  
and was treated in the same manner.
The case of the contamination due to the $\beta$-delayed neutron branch
is more challenging. The decay simulation must explicitly include the emitted neutrons.
These neutrons interact with detector materials producing $\gamma$ rays through inelastic
and capture processes.
An event generator was implemented which reproduces the known
neutron energy distribution, taken from~\cite{endfb71}, and the known $\gamma$-ray
intensity in the final nucleus, taken from~\cite{br87,br88,rb94}. The event generator 
requires the $\beta$ intensity distribution followed by
neutron emission $I_{\beta n}$ 
which was obtained from deconvolution of the neutron spectrum.
Another issue is whether the interaction
of neutrons with the detector can be simulated accurately. 
We have shown recently~\cite{tai15} that this is indeed the case
provided that Geant4 is updated  
with the newest neutron data libraries and the original capture cascade 
generator is substituted by an improved one.
The normalization factor of the $\beta$-delayed neutron decay 
contamination is fixed by the $P_{n}$ value.
Another important source of spectrum distortion is the
summing-pileup of events. 
If more than one event arrives within the same ADC event gate, a signal with 
the wrong energy is stored in the spectrum.
Apart from the electronic pulse pile-up effect for a single detector module~\cite{can99} 
one must 
consider the summing of signals from different detector modules. 
A new Monte Carlo (MC) procedure to calculate their combined contribution has been developed. 
The procedure is based on the random superposition of two stored events 
within the ADC gate length. 
The normalization of the resulting summing-pileup spectrum is fixed by the 
event rate and the ADC gate length~\cite{can99}.

%%%%%  Figure 1  %%%%%
\begin{figure}[h]
 \begin{center}
 \includegraphics[width=8.6cm]{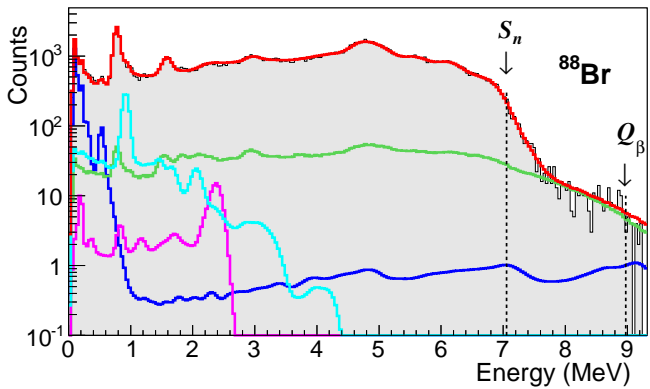}
 \caption{Relevant histograms for $^{88}$Br: parent decay 
(grey filled), daughter decay (pink), summing-pileup (green), 
$\beta$-delayed neutron decay (dark blue),
accidental contamination (light blue),  
reconstructed spectrum (red).}
 \label{fig1}
 \end{center}
\end{figure}

Several laboratory $\gamma$-ray sources were used 
to determine the energy and resolution calibration of the spectrometer. 
The measured singles spectra also served to verify the accuracy of the 
spectrometer response simulated with Geant4. 
The use of $\beta$-gated spectra in the analysis required additional verifications
of the simulation. Due to the existence of an electronic threshold in the Si detector (100~keV)
the $\beta$-detection
efficiency has a strong dependence with $\beta$-endpoint energy up to about 2~MeV.
This affects the region of interest (see Fig.~\ref{fig1}).
To verify that the MC simulation reproduces this
energy dependence we use the information
from a separate experiment~\cite{agr14} measuring $P_{n}$ values with the
neutron counter BELEN and
the same $\beta$ detector. Several
isotopes with 
different neutron emission windows $Q_{\beta}- S_{n}$
were measured, resulting in variations of the neutron-gated $\beta$ efficiency
as large as 25\%.
Geant4 simulations
using the above mentioned $\beta$-delayed neutron decay generator
are able to reproduce the isotope-dependent efficiency within better than 4\%.

Figure~\ref{fig1} shows the $\beta$-gated TAGS spectrum measured during the
implantation of $^{88}$Br ions.
Also shown is the contribution  of the daughter $^{88}$Kr decay, the neutron decay branch 
populating $^{87}$Kr, the summing-pileup contribution and the 
accidental contamination of $^{94}$Y.  
Notice the presence of net counts beyond the
neutron separation energy, which can only be attributed to the decay feeding excited states 
above $S_{n}$ which de-excite by $\gamma$-ray emission. In this region
the major background contribution comes from summing-pileup which is well reproduced
by the calculation as can be observed. 
Similar pictures were obtained for 
the decay of $^{87}$Br and $^{94}$Rb.

The analysis of the $\beta$-gated spectra follows the method developed by the Valencia 
group~\cite{tai07a,tai07b}. The intensity distribution $I_{\beta \gamma}$ 
is obtained by deconvolution 
of the TAGS spectrum with the calculated spectrometer response to the decay. 
The response to electromagnetic
cascades is calculated from a set of branching ratios (BR) and the MC calculated 
response to individual $\gamma$ rays.
Branching ratios are taken from~\cite{br87,br88,rb94} for the low energy part 
of the decay level scheme.
The excitation energy range above the last discrete level is treated as a
continuum
divided into 40~keV bins. Average 
BR for each bin are calculated from NLD and PSF 
as prescribed by the Hauser-Feshbach model. 
We use NLD from Ref.~\cite{gor08}  as
tabulated in the RIPL-3 library~\cite{ripl3}.
The PSF is obtained 
from Generalized Lorentzian (E1) or Lorentzian (M1, E2) functions using the parameters 
recommended in Ref.~\cite{ripl3}.  The electromagnetic 
response is then convoluted with the simulated 
response to the $\beta$ continuum.
The spin-parity of some of the discrete states at low excitation energy
in the daughter nucleus is uncertain. They are however required 
to calculate the BR from the states in the continuum.
The unknown spin-parities were varied and those values giving the best reproduction
of the spectrum were adopted. 
There is also ambiguity in the spin-parity 
of the parent nucleus which determines the spin-parity of the levels 
populated in the continuum. Here we assume that allowed Gamow-Teller (GT) selection rules apply.
Our choices, $3/2^{-}$ for $^{87}$Br, $1^{-}$ for $^{88}$Br and $3^{-}$ for
$^{94}$Rb, are also based on which values best reproduce the spectrum.

%%%%%  Figure 2  %%%%%
\begin{figure}[h]
 \begin{center}
 \includegraphics[width=8.6cm]{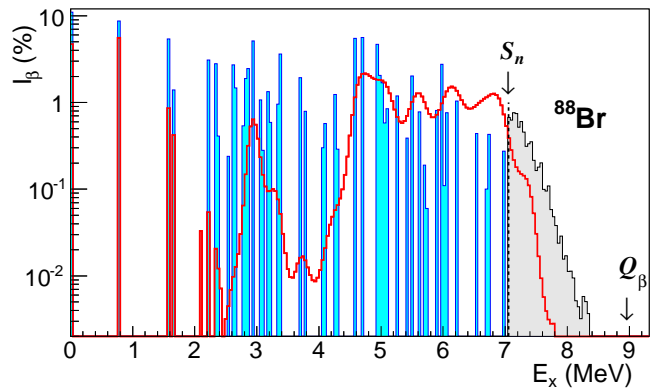}
 \caption{Beta intensity distributions for $^{88}$Br: TAGS result (red),
high-resolution $\gamma$ spectroscopy (blue), 
from $\beta$-delayed neutron (grey filled).}
 \label{fig2}
 \end{center}
\end{figure}

As an example of the results of the analysis we show  in Fig.~\ref{fig2}
the $I_{\beta \gamma}$  intensity obtained for $^{88}$Br. 
The spectrum reconstructed with this intensity
distribution reproduces well the measured spectrum (see Fig.~\ref{fig1}). 
The analysis for the other two isotopes shows
similar quality in the reproduction of the spectra.
We also include
in Fig.~\ref{fig2} the intensity obtained from high-resolution
measurements~\cite{br88}, showing a strong {\it Pandemonium} effect. 
The {\it Pandemonium} effect is even stronger in the case of $^{94}$Rb
and somewhat less for $^{87}$Br. The complete $I_{\beta \gamma}$ 
and its impact on reactor decay heat~\cite{alg10} and 
antineutrino spectrum~\cite{fal12} summation calculations
will be discussed elsewhere~\cite{val15}. 
Here we concentrate on the portion of that intensity 
located in the neutron unbound region. A sizable TAGS intensity is observed 
above $S_{n}$ extending well beyond the first few hundred keV
where the low neutron penetrability makes $\gamma$-ray emission competitive.
For comparison
Fig.~\ref{fig2} also shows 
$I_{\beta n}$ deduced from the neutron spectrum~\cite{endfb71} as explained above. 
The $I_{\beta \gamma}$ above $S_{n}$ adds up to
 $\sum I_{\beta \gamma} =1.59(17)$\%, to be compared with the integrated 
$I_{\beta  n}$  (or $P_{n}$) of 6.4(6)\%.
From the TAGS analysis for the other two isotopes we find a
$\sum I_{\beta \gamma}$  of 3.5(3)\% ($^{87}$Br) and 0.53(14)\% ($^{94}$Rb)
to be compared with $P_{n}$-values of 2.60(4)\% and 10.18(24)\% respectively. 
In the case of $^{87}$Br we find 7 times more intensity than the high-resolution 
measurement~\cite{ram83}. 
The uncertainty quoted on $\sum I_{\beta \gamma}$ is dominated by
systematic uncertainties. We did a careful evaluation of possible sources
of systematic effects for each isotope.
The uncertainty coming
from assumptions in the BR varies from 1\% to 5\%
(relative value) depending on the isotope. The impact of the use of different
deconvolution algorithms~\cite{tai07b} is in the range of 2\% to 10\%. The uncertainty
in the energy dependence of the $\beta$ efficiency contributes with 4\%.
The major source of uncertainty comes from the normalization
of the background contribution, which at the energies of interest
is dominated by the summing-pileup. We estimated that reproduction
of spectra could accommodate  at most a $\pm15$\% variation from the nominal value,
which translates into uncertainties of 6\% to 22\%.

\begin{figure}[h]
 \begin{center}
 \includegraphics[width=8.6cm]{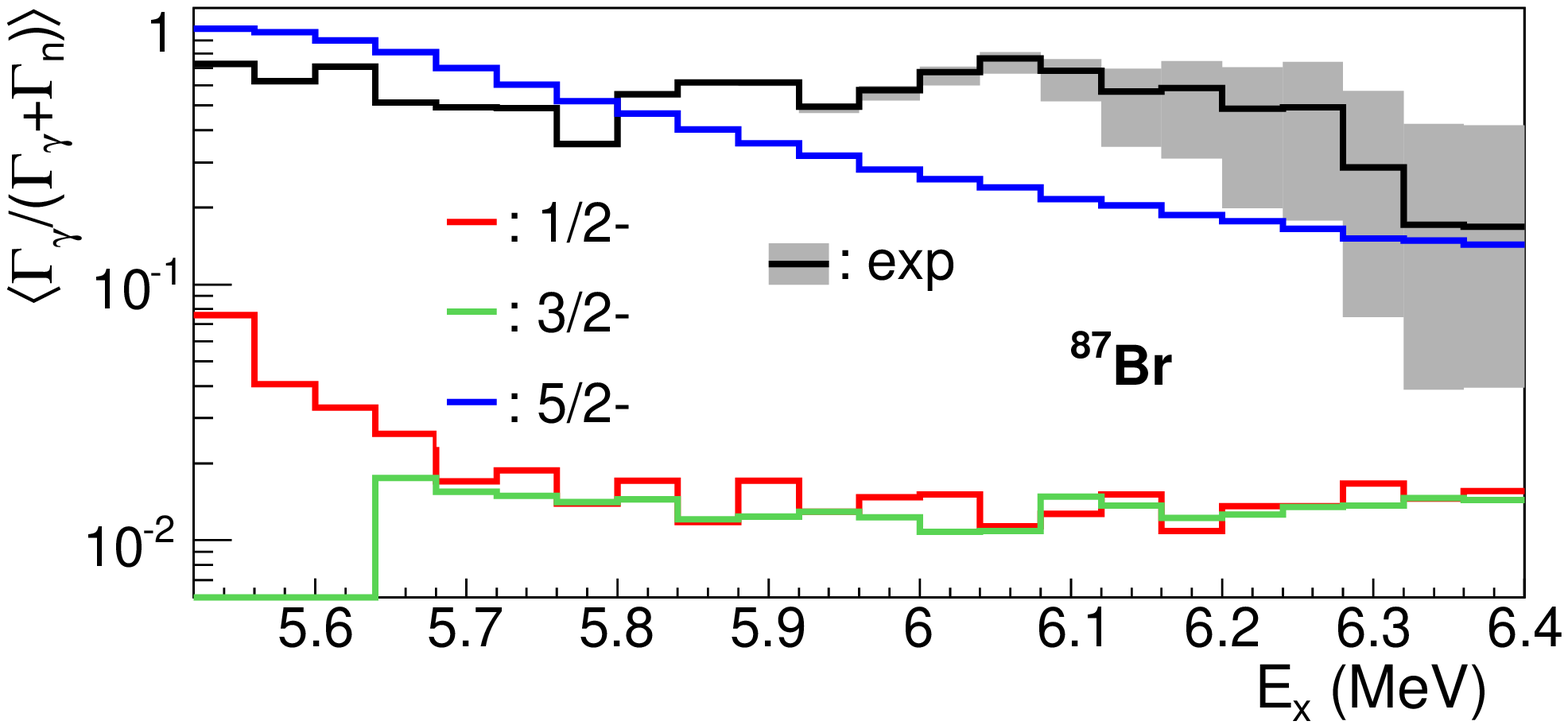}
 \includegraphics[width=8.6cm]{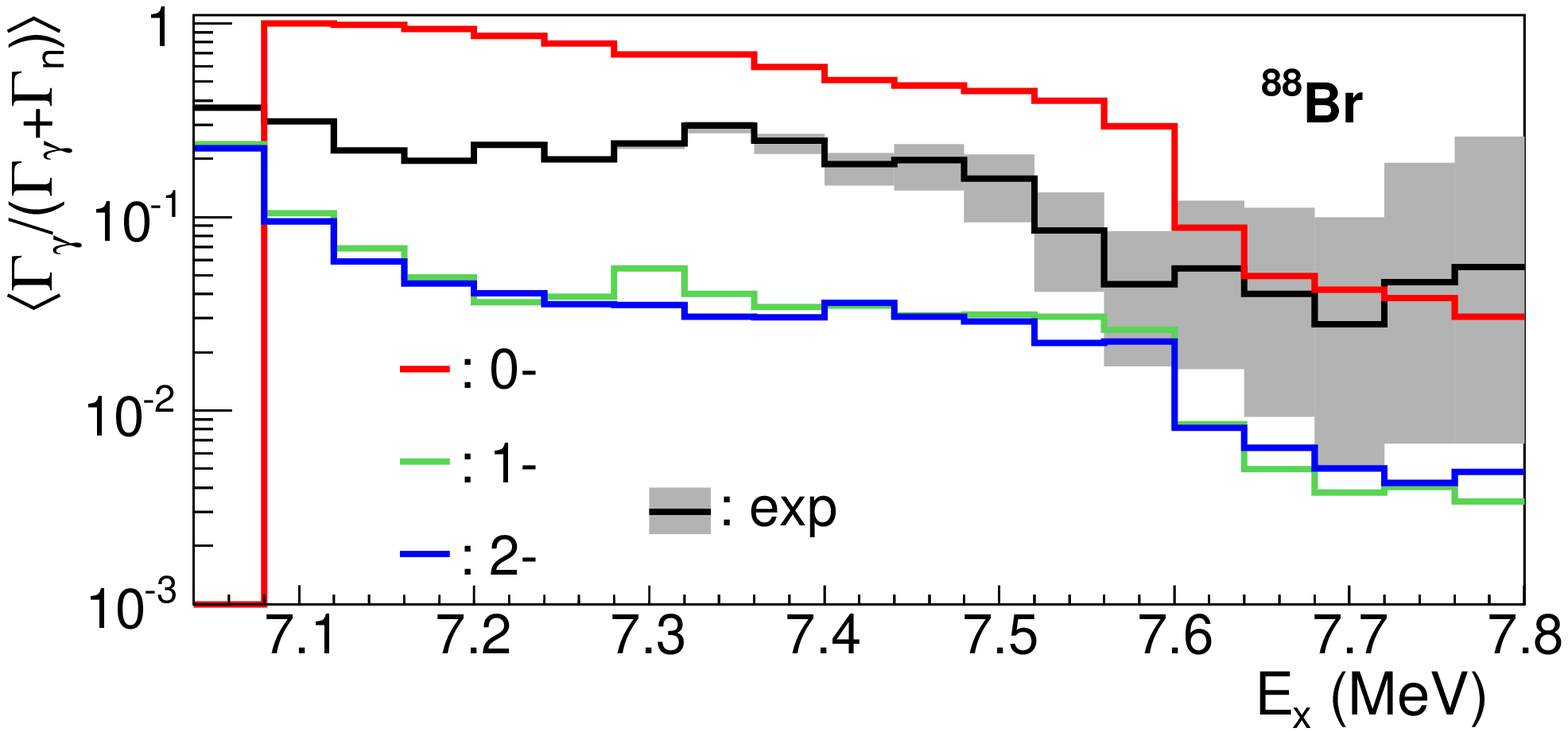}
 \includegraphics[width=8.6cm]{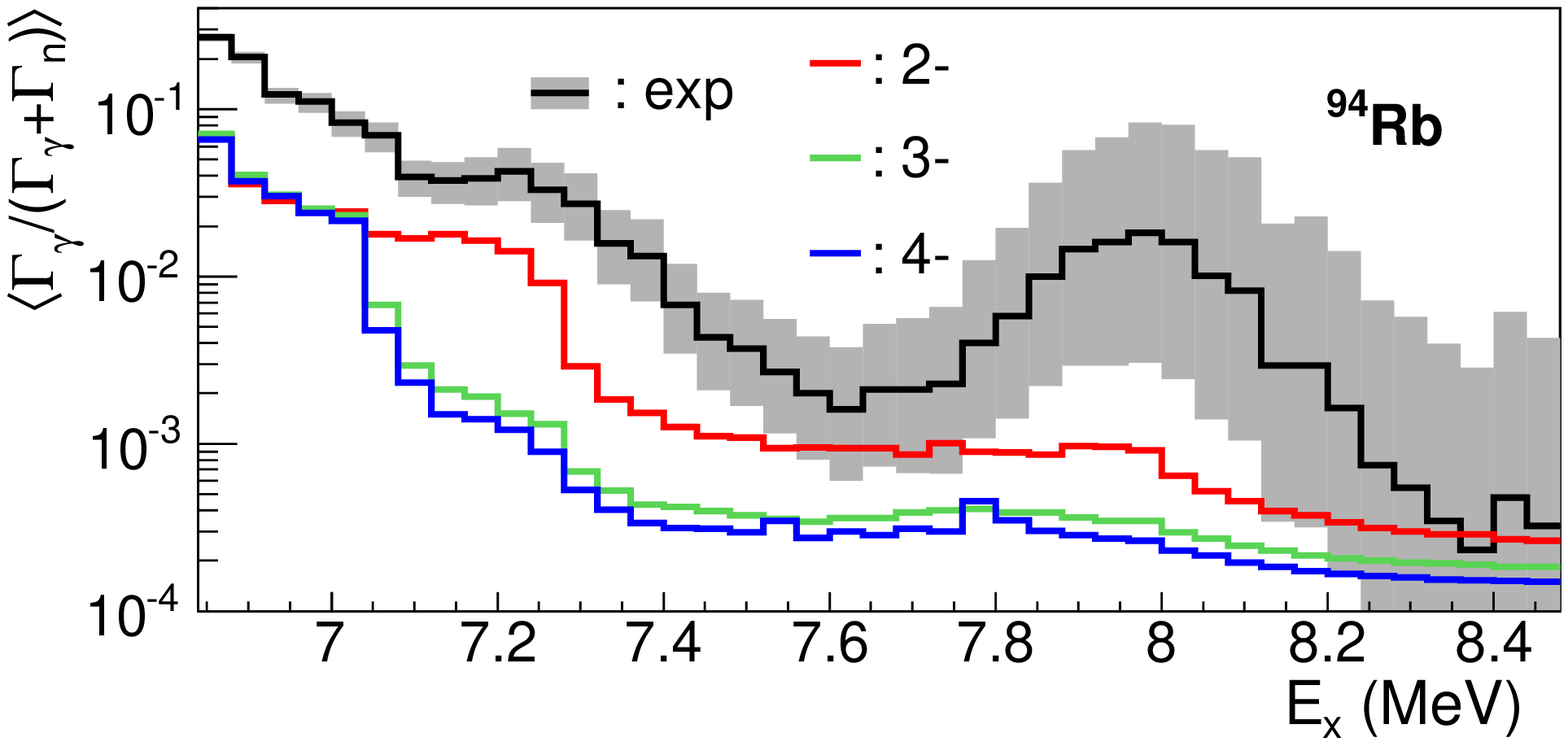}
 \caption{Average gamma to total width ratio from experiment and calculated for
   the three spin-parity groups populated in allowed decays. The 
shaded area around the experimental value indicates the sensitivity to 
the background normalization (see text).}
 \label{fig3}
 \end{center}
\end{figure}

Figure~\ref{fig3} shows the ratio $I_{\beta \gamma} / (I_{\beta \gamma} + I_{\beta  n})$
in the range of energies analyzed with TAGS
for all three cases.
This ratio is identical to the average ratio 
$\langle \Gamma_{\gamma}/(\Gamma_{\gamma} + \Gamma_{n}) \rangle$
over all levels populated in the decay. 
The shaded area around the experimental value in Fig.~\ref{fig3}
serves to indicate the sensitivity of the
TAGS results to background normalization as indicated above. 
The  average width ratio was calculated using the Hauser-Feshbach model. 
The results
for the three spin-parity groups populated in GT decay are shown.
The NLD and PSF values used in these calculations
are the same as those used  in the TAGS analysis.
The new ingredient needed is the NTC, which is obtained from the 
Optical Model (OM) with the TALYS-1.4 software package~\cite{talys}. 
OM parameters are taken from the so-called local
parametrization of Ref.~\cite{kon03}.
Neutron transmission is calculated for known final levels populated in the decay~\cite{br87,br88,rb94}.
In order to compute the average width ratio we need to include the effect of
statistical fluctuations in the individual widths~\cite{jon76}.
We use the MC method to obtain the average of width ratios.
The sampling procedure is analogous to that described in Ref.~\cite{tai07a}.
Level energies for each spin-parity are generated
according to a Wigner distribution and their corresponding 
$\Gamma_{\gamma}$ and $\Gamma_{n}$ to individual final states
are sampled from PT distributions.
The total $\gamma$ and neutron widths are obtained 
by summation over all possible final states and the ratio
computed. The ratio is averaged for all levels lying
within each energy bin. In order to suppress fluctuations  
in the calculated average, the sampling procedure is repeated
between 5 and 1000 times depending on level density.
Very large average enhancement factors were obtained, reaching
two orders-of-magnitude when the neutron emission is dominated
by the transition to a single final state.

%%%%%  Figure 3  %%%%%

In the case of $^{87}$Br $3/2^{-}$ decay one can see in Fig.~\ref{fig3} that the
strong $\gamma$-ray emission above $S_{n}$ 
can be explained as a consequence of the large hindrance
of $l=3$ neutron emission from $5/2^{-}$ states in $^{87}$Kr 
to the $0^{+}$ g.s. of  $^{86}$Kr, as pointed out in Ref.~\cite{sla72}. 
In the case of $^{88}$Br $1^{-}$ decay a
similar situation occurs for $0^{-}$ states in $^{88}$Kr
below the first excited state in $^{87}$Kr at 532~keV,
which require  $l=3$ to populate the $5/2^{+}$
g.s. in $^{87}$Kr. 
For a more quantitative assessment
one should know the distribution of $\beta$ intensity
between the three spin groups,
which could be obtained from $\beta$-strength theoretical
calculations.
The case of $^{94}$Rb $3^{-}$ decay is
the most interesting. 
The final nucleus $^{93}$Sr 
is five neutrons away from $\beta$ stability.
The $\gamma$ intensity although strongly reduced,
only 5\% of the neutron intensity, is detectable 
up to 1.5~MeV beyond $S_{n}$. The structure observed
in the average width ratio,
is associated
with the opening of $\beta$n channels to different
excited states.
Note that the structure is reproduced by the calculation, 
which confirms the energy calibration at high excitation energies.
In any case the
calculated average gamma-to-total ratio is well
below the experiment. 
In order to bring the calculation
to the experimental value one would need to enhance
the PSF, or suppress the NTC, or any suitable combination of the two, 
by a very large factor.
For instance we verified that a twenty-fold increase of the E1 PSF
would reproduce the measurement assuming a $\beta$-intensity
spin distribution proportional to $2J+1$.
An enhancement
of such magnitude for neutron-rich nuclei,
leading to a similar enhancement of $(\mathrm{n},\gamma)$ cross sections,
will likely have an impact on $r$-process abundance calculations.
Therefore it will be important to investigate the magnitude of possible variations of the NTC.

In conclusion, we have confirmed the suitability of the TAGS technique to obtain
accurate information on $\gamma$-ray emission from neutron unbound states 
and applied it to three known $\beta$-delayed neutron emitters.
A surprisingly large $\gamma$-ray branching of  57\% and 20\%  
was observed for  $^{87}$Br and  $^{88}$Br respectively,
which can be explained as a nuclear structure effect.
In the case of $^{87}$Br
we observe 7 times more intensity than previously detected with high resolution $\gamma$-ray
spectroscopy, which confirms the need of the TAGS technique for such studies.
In the case of the more neutron-rich $^{94}$Rb
the measured branching is only 4.5\% but still much larger than
the results of Hauser-Feshbach statistical calculations,
after proper correction for individual width fluctuations.
The large difference between experiment and calculation 
can be reconciled by an enhancement of standard PSF of over one order-of-magnitude.
To draw more general conclusions it will be necessary to extend this type of study to other
neutron-rich $\beta$-delayed neutron emitters.
Such measurements using the TAGS technique are already underway and 
additional ones are planned.

This work was supported by Spanish Ministerio de Econom\'{\i}a y
Competitividad under grants FPA2008-06419, FPA2010-17142 and FPA2011-
24553, CPAN CSD-2007-00042 (Ingenio2010), and by EPSRC and STFC (UK).
Work at ANL was
supported by the U.S Department of Energy, Office of Science, Office of Nuclear
Physics, under contract number DE-AC02-06CH11357.


\end{document}